\documentclass[11pt]{article}

\usepackage{authblk}

\usepackage{float}
\usepackage{array}

\usepackage{tikz}
\usetikzlibrary{positioning,fit,calc,arrows.meta}

\tikzset{>=Latex}

\usepackage{tabularx}

\usepackage{graphicx}
\renewcommand{\arraystretch}{1.1}

\usepackage[margin=1in]{geometry}
\usepackage{authblk}
\usepackage{amsmath, amssymb, amsthm}
\usepackage{bm}
\usepackage{microtype}
\usepackage{booktabs}
\usepackage{array}
\usepackage{multirow}
\usepackage{graphicx}
\usepackage{tikz}

\usepackage{algorithm}
\usepackage{algpseudocode}
\usepackage{listings}
\usepackage{xcolor}
\usepackage[numbers,sort&compress]{natbib}
\usepackage{hyperref}
\hypersetup{
  colorlinks=true,
  linkcolor=blue,
  citecolor=blue,
  urlcolor=blue
}

\title{Execution Is the New Attack Surface:\\
Survivability-Aware Agentic Crypto Trading with OpenClaw-Style Local Executors}

\author{

\includegraphics[height=2cm]{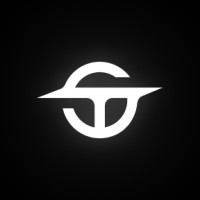}\\
True Trading -- AI\\
\vspace{1.2em} 

\begin{center}
\setlength{\tabcolsep}{0pt} 
\begin{minipage}[t]{0.28\textwidth}\centering
\textbf{Ailiya Borjigin}\\
True Trading\\
\texttt{ailiya.borjigin@gmail.com}
\end{minipage}\hspace{12pt}%
\begin{minipage}[t]{0.22\textwidth}\centering
\textbf{Igor Stadnyk}\\
True Trading\\
\texttt{igor@true.trading}
\end{minipage}\hspace{12pt}%
\begin{minipage}[t]{0.22\textwidth}\centering
\textbf{Ben Bilski}\\
True Trading\\
\texttt{ben@true.trading}
\end{minipage}

\vspace{1.2em} 

\begin{minipage}[t]{0.26\textwidth}\centering
\textbf{Serhii Hovorov}\\
Inc4.net\\
\texttt{s.hovorov@inc4.net}
\end{minipage}\hspace{18pt}%
\begin{minipage}[t]{0.26\textwidth}\centering
\textbf{Sofiia Pidturkina}\\
Inc4.net\\
\texttt{s.pidturkina@inc4.net}
\end{minipage}
\end{center}

}

\date{}

\begin{document}
\maketitle
\begin{abstract}
\textbf{OpenClaw-style} agent stacks turn language into \emph{privileged execution} by routing LLM intents through tool interception, policy gateways, and a local executor.
In parallel, skill marketplaces such as \textbf{skills.sh} operationalize capability acquisition via installable skills and CLIs, creating a fast-growing \emph{capability supply chain}.
Together, these trends shift the dominant safety failure mode from ``wrong answers'' to \emph{execution-induced loss}: untrusted prompts, compromised skills, or narrative manipulation can translate into real trades and irreversible side effects.

We propose \textbf{Survivability-Aware Execution (SAE)}, an execution-layer survivability standard designed for OpenClaw-style systems and skill-enabled agents.
SAE is deployed as middleware between any strategy engine (LLM or non-LLM) and the exchange executor.
It implements an explicit execution contract (\texttt{ExecutionRequest}, \texttt{ExecutionContext}, \texttt{ExecutionDecision}) and enforces non-bypassable invariants at the last mile: projection-based exposure budgeting, cooldown and order-rate limits, slippage bounds, staged execution, and tool/venue allowlists.
To make delegated execution empirically testable under skill supply-chain risk, we operationalize the \textbf{Delegation Gap (DG)} using a logged Intended Policy Spec that yields deterministic out-of-scope labeling and reproducible DG metrics.

On a reproducible offline replay built from official Binance USD-M BTCUSDT/ETHUSDT perpetual data (15m; 2025-09-01--2025-12-01, including funding),
SAE substantially improves survivability and robustness.
Relative to \textsc{NoSAE}, SAE reduces maximum drawdown from $0.4643$ to $0.0319$ (Full; $93.1\%$), shrinks tail-loss magnitude $|\mathrm{CVaR}_{0.99}|$ from $4.025\times10^{-3}$ to $\approx 1.02\times10^{-4}$ ($\sim 97.5\%$),
and lowers DG loss proxy from $0.647$ to $0.019$ ($\sim 97.0\%$), while reducing AttackSuccess from $1.00$ to $0.728$ with zero FalseBlock in this run.
Dependence-aware tests (block bootstrap, paired Wilcoxon, two-proportion test) confirm the shifts are statistically detectable.
Overall, SAE reframes agentic trading safety for the OpenClaw+skills era: treat upstream intent and skills as untrusted, and enforce survivability where actions become side effects.
\end{abstract}

\section{Introduction and Real-world Motivation for SAE}
\label{sec:intro}

Tool-using agents have become a mainstream engineering pattern. \citet{yao2022react} demonstrates interleaving reasoning with actions that consult external sources, and \citet{schick2023toolformer} shows that models can be trained (self-supervised) to decide \emph{when} and \emph{how} to call tools. As these ideas move into production, the safety boundary shifts: language is no longer only an output medium, but a \emph{control surface} for real-world side effects. The key question is therefore not only whether an agent produces correct text, but whether its \emph{execution interface} converts untrusted intent into harmful actions.

\paragraph{Why execution becomes the primary attack surface.}
This execution boundary is made explicit by \emph{OpenClaw-style} agent stacks. OpenClaw separates (i) where tools run (sandbox vs.\ host), (ii) which tools are available (tool policy), and (iii) an explicit escape hatch for host execution under elevated settings \citep{openclawSandboxVsToolPolicyElevated}. Its security guidance recommends strict allowlists for high-risk tools, sandboxing whenever untrusted inputs are in the loop, and keeping secrets out of prompts and the agent-accessible filesystem \citep{openclawSecurity}. These design choices capture a security principle that motivates our paper: when tool access is permissive, untrusted text can become privileged actions. In finance, the consequences are immediate because side effects are monetized.

\paragraph{Why the skills ecosystem amplifies the problem.}
A second trend makes the problem harder: \emph{installable skill ecosystems} that operationalize capability acquisition. The \emph{Agent Skills} standard packages reusable capability context as directories anchored by \texttt{SKILL.md} with YAML frontmatter and progressive-disclosure design \citep{xu2026agentskills}. Vercel’s open skills ecosystem provides a CLI workflow (e.g., \texttt{npx skills add \dots}) and a discovery surface via \texttt{skills.sh} \citep{vercel2026skills,vercelSkillsCLI}. In parallel, OpenClaw positions \emph{ClawHub} as a public registry for discovering, installing, updating, and syncing skills \citep{openclawClawHub,openclawSkills}. Together, OpenClaw-style privileged execution and skills.sh-style marketplaces imply a near-term reality: capabilities will be frequently acquired from third parties, and the \emph{capability supply chain} becomes part of the attack surface. This portability is valuable for productivity, but it increases risk because installing a third-party skill is operationally close to importing code and instructions that can influence tool calls and execution parameters. In early 2026, security reports documented malware distributed via marketplace-hosted OpenClaw skills, including crypto-themed skills designed to harvest credentials and other sensitive data \citep{roth2026openclawSkillsNightmare,james2026maliciousOpenclawSkills}. These incidents support a conservative stance: treat skill installation as a high-risk event and assume adversarial incentives exist wherever execution privileges intersect with asset custody.

\paragraph{Why ``assume compromise'' is a reasonable baseline.}
Agent networks can further amplify risk through connectivity. Reuters reported that Moltbook---a Reddit-like social site marketed for AI agents---suffered a major exposure due to a basic security flaw, leaking private agent messages, thousands of owner emails, and a large volume of credentials \citep{reuters2026moltbookHole}. Even when downstream impact differs across deployments, the lesson is structural: fast-growing agent ecosystems can externalize sensitive data at scale through simple misconfigurations. When agents are both highly connected and highly privileged, ``assume compromise'' becomes a reasonable baseline. This stance aligns with broader directions in AI-governed, web-trustworthy architectures that treat governance and trust signals as first-class system inputs \citep{borjigin2025aigoverned}.

\paragraph{Why crypto perpetuals amplify execution mistakes.}
Crypto perpetual trading is a particularly sharp setting for execution-layer safety. Binance documentation notes that funding fees are deducted from a trader's futures wallet balance and, if insufficient, from position margin---potentially shifting liquidation price and increasing liquidation risk \citep{binanceFundingRates}. Binance also documents that maintenance margin requirements depend on notional tiers and directly affect liquidation thresholds \citep{binanceLiquidationProtocols,binanceLeverageMarginTiers}. As a result, execution choices that may look minor in an intent string---such as leverage, order rate, slippage tolerance, and timing under stress---can be structurally amplified into nonlinear tail outcomes. Relatedly, constrained formulations of trade execution and independent audit layers have been proposed as system designs to enforce hard participation and compliance constraints, complementing execution-layer enforcement such as SAE \citep{borjigin2025safecompliant}.

\paragraph{Survivability-Aware Execution (SAE).}
These conditions motivate \textbf{Survivability-Aware Execution (SAE)}: a standardized execution-layer safety contract designed to plug into OpenClaw-style tool interception and remain robust under skills.sh-style capability supply chains. SAE is designed to be strategy-agnostic and composable within broader agentic trading architectures \citep{borjigin2025decentralizedtrading}. SAE treats upstream outputs as \emph{untrusted intent} and enforces survivability constraints as \emph{code-level policies} (budgets, cooldowns, tool gating, and kill-switches) before privileged actions can occur.

\paragraph{Contributions.}
Concretely, our main contributions are: (i) an operational DG measurement protocol grounded in Intended Policy Specs and hard out-of-scope rules; (ii) a survivability-first execution contract compatible with OpenClaw-style tool interception and skills.sh-style capability ecosystems; (iii) practical enforcement algorithms with an exposure-based theoretical bound linking projection to worst-case loss amplification; and (iv) a fully reproducible Binance replay evaluation reporting survivability, robustness, and overhead metrics across SAE variants.

\paragraph{Paper roadmap.}
We proceed as follows. Section 2.1 formalizes the agentic execution setting and defines the Delegation Gap (DG) using a logged Intended Policy Spec, enabling deterministic out-of-scope labeling and reproducible DG metrics. Section~\ref{sec:sae_design} specifies SAE as an OpenClaw-style execution contract with non-bypassable enforcement (projection-based budgeting, temporal guards, and trust-conditioned tightening under skill supply-chain risk). Sections~\ref{sec:eval_binance}--\ref{sec:eval_results} evaluate SAE on a reproducible Binance USD-M replay with attack instrumentation and dependence-aware statistical tests, and Sections~\ref{sec:discussion}--\ref{sec:limitations} discuss trade-offs and limitations.

\section{Problem Formulation}

\subsection{Formal Delegation Gap}
\label{sec:dg}

Let $\mathcal{A}_{\text{intended}}$ be the set of actions the operator intends to authorize in the current context,
and let $\mathcal{A}_{\text{actual}}$ be the set of actions the deployed system can execute given its tools,
permissions, credentials, and integrations (including indirect effects from skills/plugins and elevated local execution).
In agentic execution it is typical that $\mathcal{A}_{\text{intended}} \subsetneq \mathcal{A}_{\text{actual}}$,
and the effective support of executed actions can expand under prompt injection, compromised inputs, or supply-chain events.

\paragraph{Delegation Gap (DG).}
We define the Delegation Gap as the expected loss introduced by actions that are executable but outside intended scope:
\begin{equation}
\mathrm{DG} \;\triangleq\; \mathbb{E}\big[\, \ell(a_t)\cdot \mathbf{1}\{a_t \notin \mathcal{A}_{\text{intended}}(S_t)\}\,\big],
\quad a_t \sim P_{\pi}(\cdot)\ \text{over}\ \mathcal{A}_{\text{actual}} ,
\label{eq:dg_def}
\end{equation}
where $\ell(\cdot)$ is a fixed loss functional (e.g., realized PnL loss, liquidation indicator/proxy, or security loss proxy),
and $S_t$ is an explicit \emph{Intended Policy Spec} (defined next).
DG is measurable in a reproducible way because it reduces to (i) a hard out-of-scope test, plus (ii) a fixed loss proxy.

\subsubsection{Operationalizing $\mathcal{A}_{\text{intended}}$ via an Intended Policy Spec}
\label{sec:aintended_spec}

To avoid treating ``user natural language'' as the ground truth of intent, we define intent as a structured specification:
\begin{equation}
S_t \;=\; (T_t, R_t, M_t, U_t),
\label{eq:intended_spec}
\end{equation}
with the following components.

\paragraph{(i) Allowed action/tool set $T_t$.}
A finite set of permitted action types and tools, e.g.,
$\{\texttt{open},\texttt{close},\\ \texttt{modify},\texttt{cancel}\}$ plus permitted venues/symbols/accounts.

\paragraph{(ii) Risk budgets $R_t$.}
Hard caps on execution-relevant exposures, e.g., maximum leverage, maximum notional, maximum order rate,
maximum slippage bound, maximum holding time, maximum concurrent positions.

\paragraph{(iii) Market-state constraints $M_t$.}
State-triggered tightening rules based on market regime indicators (e.g., volatility/liquidity/funding extremes),
so that high-risk regimes reduce feasible budgets.

\paragraph{(iv) User/account constraints $U_t$.}
Account-state constraints such as margin ratio, drawdown/PnL state, cooldown timers, and exposure concentration.

\paragraph{Induced intended action set.}
We then define:
\begin{equation}
\mathcal{A}_{\text{intended}}(S_t)
\;=\;
\{\, a \;:\; a \text{ satisfies } T_t, R_t, M_t, U_t \,\}.
\label{eq:aintended}
\end{equation}

\subsubsection{Out-of-scope rules (hard, reproducible)}
\label{sec:oos_rules}

An executed/attempted action $a_t$ is labeled \emph{out-of-scope} if it violates any of the following rule classes:

\begin{itemize}
\item \textbf{Cap violation (risk-budget breach):}
exceeds leverage/notional/order-rate/slippage/holding-time caps specified by $R_t$ (possibly tightened by $M_t$ and $U_t$).
\item \textbf{Tool/venue violation (capability breach):}
invokes an unauthorized tool, market, venue, symbol universe, or cross-account capability not allowed by $T_t$.
\item \textbf{State violation (context breach):}
executes actions that are disallowed under current market/account states (e.g., ``extreme volatility'' or ``low margin ratio''
states where only reduce-only actions are permitted).
\end{itemize}

\subsubsection{DG estimators and reporting}
\label{sec:dg_estimators}

Given logs of attempted/executed actions $\{a_t\}_{t=1}^{N}$, we report:

\paragraph{Out-of-scope rate.}
\begin{equation}
\widehat{\mathrm{DG}}_{\text{rate}}
\;=\;
\frac{1}{N}\sum_{t=1}^{N}\mathbf{1}\{a_t \notin \mathcal{A}_{\text{intended}}(S_t)\}.
\label{eq:dg_rate}
\end{equation}

\paragraph{Out-of-scope loss contribution (proxy).}
For a fixed, pre-declared loss proxy $\ell(\cdot)$:
\begin{equation}
\widehat{\mathrm{DG}}_{\text{loss}}
\;=\;
\frac{\sum_{t=1}^{N}\ell(a_t)\cdot \mathbf{1}\{a_t \notin \mathcal{A}_{\text{intended}}(S_t)\}}
{\sum_{t=1}^{N}\big|\ell(a_t)\big| + \epsilon},
\label{eq:dg_loss}
\end{equation}
where $\epsilon$ is a small constant to avoid division by zero.

\paragraph{Paper-facing DG table.}
In experiments we report (i) $\widehat{\mathrm{DG}}_{\text{rate}}$, (ii) $\widehat{\mathrm{DG}}_{\text{loss}}$,
and (iii) the reduction under SAE vs.\ No-SAE. See Table 3.

\subsection{Survivability-first objective as constrained optimization}
Let $\pi$ denote the composite policy: strategy $\rightarrow$ SAE gating $\rightarrow$ execution runtime.
Let $R_{\pi}$ be the return random variable over a horizon (including funding cash flows).
SAE targets tail-risk reduction under minimal performance constraints:
\begin{equation}
\label{eq:cvar_opt}
\min_{\pi \in \Pi}\ \mathrm{CVaR}_{\alpha}(R_{\pi})
\quad \text{s.t.}\quad
\mathbb{E}[R_{\pi}] \ge \mu_0,\quad
\Pr(\mathrm{Liquidation}_{\pi}) \le \epsilon.
\end{equation}
CVaR (expected shortfall) is a standard tail-risk objective \citep{rockafellar2000cvar,rockafellar2002cvar}.
The key point is conceptual: SAE is not ``an alpha clamp''; it is a constraint layer that alters the feasible policy set by restricting execution.

\paragraph{From conceptual objective to reproducible selection.}
In practice, $\pi$ is instantiated by a parameterized strategy and a parameterized SAE gate.
We write $\pi=\pi_{\phi,\theta}$ where $\phi \in \Phi$ are strategy parameters (e.g., lookbacks, thresholds)
and $\theta \in \Theta$ are SAE gate parameters (e.g., budgets, cooldown, trust-conditioned tightening, staged enforcement).
Given a replay dataset $\mathcal{D}$ (synthetic or exchange historical data), we evaluate $\pi_{\phi,\theta}$ via a deterministic simulator
$\mathcal{E}(\phi,\theta;\mathcal{D})$ that returns a metrics vector including drawdown, tail losses, security, and usability outcomes.

\paragraph{Operational constraints for agentic execution.}
To make the feasibility set explicit for tool-using agents, we augment the survivability constraints in~\eqref{eq:cvar_opt}
with security \& usability constraints that are unique to delegated execution:
(i) \emph{attack success} $\mathrm{AS}(\pi)$, the fraction of injected out-of-scope attack attempts that are not blocked;
(ii) \emph{false block} $\mathrm{FB}(\pi)$, the fraction of legitimate in-scope requests that are blocked; and
(iii) optional executor overhead $\mathrm{Lat}(\pi)$.
These yield the constrained policy family
\begin{equation}
\label{eq:sae_constraints}
\Pi_{\mathrm{SAE}}(\alpha,\beta,\tau)
=\left\{
\pi \in \Pi:
\mathrm{AS}(\pi)\le \alpha,\ 
\mathrm{FB}(\pi)\le \beta,\ 
\mathrm{Lat}(\pi)\le \tau
\right\},
\end{equation}
where $\alpha$ controls acceptable adversarial leakage, $\beta$ bounds usability degradation, and $\tau$ (optional) caps last-mile latency.

\paragraph{Best-so-far constrained search (optimization protocol).}
Equation~\eqref{eq:cvar_opt} is simulator-defined, non-convex, and typically non-differentiable in $(\phi,\theta)$.
We therefore instantiate a black-box \emph{constrained} search to obtain a reproducible \emph{best-so-far} solution.
Concretely, we repeatedly sample candidate configurations $(\phi,\theta)$, replay them on a validation segment, and keep an incumbent
$(\phi^{(best)},\theta^{(best)})$.
A candidate is \emph{feasible} if it satisfies the constraints in~\eqref{eq:sae_constraints} (and survivability constraints such as $\Pr(\mathrm{Liquidation}_{\pi})\le \epsilon$).
Among feasible candidates, we update the incumbent if it improves a fixed selection score that proxies the tail-risk objective while penalizing execution risks:
\begin{equation}
\label{eq:best_so_far_update}
(\phi^{(best)},\theta^{(best)}) \leftarrow (\phi,\theta)
\quad\text{iff}\quad
\pi_{\phi,\theta}\in \Pi_{\mathrm{SAE}}(\alpha,\beta,\tau)
\ \wedge\ 
J(\phi,\theta) < J(\phi^{(best)},\theta^{(best)}),
\end{equation}
with
\begin{equation}
\label{eq:selection_score}
J(\phi,\theta)
=
w_{1}\,\mathrm{MDD}(\pi_{\phi,\theta})
\;+\; w_{2}\,\big|\mathrm{CVaR}_{0.99}(R_{\pi_{\phi,\theta}})\big|
\;+\; w_{3}\,\mathrm{DG\_loss}(\pi_{\phi,\theta})
\;+\; w_{4}\,\mathrm{Lat}(\pi_{\phi,\theta}),
\end{equation}
where $\mathrm{MDD}$ and $\mathrm{CVaR}_{0.99}$ are computed on replay returns (including funding),
$\mathrm{DG\_loss}$ is a fixed delegation-gap cost proxy under an intended policy specification,
and $w_i$ are user-chosen weights.
We run the search in batches and terminate by a compute budget or an early-stopping criterion (e.g., stop after $K$ consecutive batches with no incumbent improvement),
which yields a reproducible best-so-far configuration and a complete search trace.

\paragraph{Selection vs.\ reporting.}
To avoid post-hoc ``cherry-picking,'' the constrained search is performed on a validation segment (or inner fold),
then the chosen $(\phi^\star,\theta^\star)$ is frozen and evaluated once on a disjoint test segment (or outer fold),
with uncertainty estimated via multiple seeds where applicable.
This protocol is implemented in our released code and summarized in the walk-forward tuning procedure (Sec.5.4).

\subsection{SAE execution API spec as middleware contract}
SAE’s engineering contribution is an explicit \emph{execution contract} that can be adopted by
frameworks like OpenClaw (via tool interception) or by any trading bot. The contract standardizes
the last-mile boundary between \emph{strategy intent} and \emph{exchange submission}:

\begin{itemize}
\item \textbf{ExecutionRequest:} \texttt{symbol}, \texttt{venue}, \texttt{timestamp\_ms}, \texttt{intent}, \texttt{side},
\texttt{requested\_notional}, \texttt{requested\_leverage}, \texttt{order\_type}, \texttt{max\_slippage\_bps}, \texttt{strategy\_id}.
\item \textbf{ExecutionContext:} \emph{account state} (equity, drawdown, positions, recent trades, margin ratio),
\emph{market state} (realized volatility, funding, liquidity proxy, regime label),
and a first-class \emph{trust state} $z_t$ (defined below), with an optional narrative proxy.
\item \textbf{ExecutionDecision:} \texttt{decision} $\in \{\texttt{ALLOW},\texttt{LIMIT},\texttt{BLOCK}\}$ plus enforced constraints:
leverage cap, notional cap, order-rate cap, slippage cap, cooldown, staging plan, and structured audit fields.
\end{itemize}

\subsubsection{SAE subsumes OMS risk limits and adds three agentic layers}
\label{sec:oms_plus}

A natural question is how SAE relates to conventional OMS risk checks such as leverage caps,
position-size limits, stop-losses, rate limits, and kill-switches.
Our position is not that SAE replaces OMS, but that SAE \emph{includes a standard OMS baseline}
and then adds three layers that become necessary in agentic, skills-enabled execution.

\paragraph{Layer 0: Static OMS baseline.}
At minimum, SAE can be configured to behave like a conventional OMS:
fixed leverage and position-size limits, stop-loss / reduce-only triggers, and rate limits.
This corresponds to a static control layer that assumes upstream strategy intent is trusted
and focuses on market and margin safety.

\paragraph{Layer 1: Trust-state conditioned budgeting.}
Agentic stacks acquire capabilities via installable skills and toolchains, so upstream intent is structurally untrusted.
SAE therefore conditions budgets on a first-class trust state $z_t$ (provenance, capability risk, injection alerts),
tightening exposure when trust degrades.

\paragraph{Layer 2: Supply-chain and scope enforcement.}
In skills-enabled systems, ``what the agent is allowed to do'' must be explicit and auditable.
SAE operationalizes intended scope via an Intended Policy Spec and hard out-of-scope rules,
enabling non-bypassable allowlists over tools, venues, symbols, and accounts.

\paragraph{Layer 3: Attack-aware evaluation.}
SAE treats prompt, skill supply-chain, and narrative attacks as first-class inputs and reports system metrics such as
AttackSuccess, FalseBlock, latency, DG rate, and DG loss, rather than only trading PnL metrics.
This makes robustness claims reproducible and comparable across implementations.

\paragraph{Why \textsc{NoSAE} is a realistic baseline.}
While human discretionary traders often rely on stop-losses and broker-side controls,
many deployed agentic bots connect directly from strategy logic to the executor with no intermediate gating layer.
In such systems, the strategy output is effectively the action.
We therefore use \textsc{NoSAE} to represent a common default deployment pattern:
direct strategy-to-executor wiring without a policy-intercepting middleware.
This baseline is included not to advocate unsafe practice, but to quantify how much survivability is gained
once an explicit last-mile execution contract is introduced.

\subsection{Projection-based enforcement}
We generalize ``clamping'' as a projection of a requested action $a_{\text{req}}$ into a feasible budget region $\mathcal{F}(B)$:
\begin{equation}
\label{eq:projection}
a_{\text{SAE}} \;=\; \arg\min_{a \in \mathcal{F}(B)} D(a, a_{\text{req}}),
\end{equation}
where $D(\cdot,\cdot)$ is an execution-distance functional (e.g., weighted $\ell_2$ over leverage, notional, and order rate).
When $D$ is convex and $\mathcal{F}(B)$ is convex, \eqref{eq:projection} is a convex projection.

A core engineering special case is leverage projection:
\begin{equation}
\label{eq:leverage_clamp}
L_{\text{eff}} \;=\; \min(L_{\text{req}}, L_{\text{cap}}),
\end{equation}
which is the projection of requested leverage onto $[0,L_{\text{cap}}]$.

\paragraph{Proposition (budgeted exposure yields a worst-case one-step loss bound).}
Let $e_t \in \mathbb{R}^d$ denote the \emph{exposure vector} induced by an execution action at time $t$
(e.g., components may encode notional exposure, effective leverage, order-rate, and slippage budget).
Let the SAE enforcement runtime guarantee a budget constraint
\begin{equation}
\|e_t\| \le B_t, \qquad e_t \in \mathcal{F}(B_t),
\end{equation}
where $\mathcal{F}(B_t)$ is the feasible region specified by SAE budgets (including component-wise caps)
and $\|\cdot\|$ is any chosen norm consistent with the budget semantics.

Assume the per-step loss functional $\ell_t(e)$ is $L$-Lipschitz with respect to $\|\cdot\|$, i.e.,
\begin{equation}
|\ell_t(e)-\ell_t(e')| \le L \|e-e'\|,\quad \forall e,e'.
\end{equation}
Then the worst-case additional one-step loss relative to a risk-off (zero-exposure) action is bounded by
\begin{equation}
\ell_t(e_t) - \ell_t(0) \le L\|e_t\| \le L B_t.
\end{equation}
Moreover, if SAE enforces execution via projection $e_t^{\mathrm{eff}} = \Pi_{\mathcal{F}(B_t)}(e_t^{\mathrm{req}})$,
then the loss deviation from an unbudgeted request is bounded as
\begin{equation}
\ell_t(e_t^{\mathrm{eff}}) - \ell_t(e_t^{\mathrm{req}})
\le L\|e_t^{\mathrm{eff}}-e_t^{\mathrm{req}}\|
\le L\cdot \mathrm{dist}(e_t^{\mathrm{req}},\mathcal{F}(B_t)).
\end{equation}
This statement is intentionally weak but robust: it does not assume any particular return dynamics,
and it directly links \emph{projection-based enforcement} to a deterministic reduction in worst-case
instantaneous loss amplification. Liquidation and exchange-specific margin mechanics are handled in the
empirical replay evaluation rather than in this bound.


\section{SAE Design and Algorithms}
\label{sec:sae_design}

SAE (Survivability-Aware Execution) is an \emph{execution-layer} safety and survivability middleware placed between a strategy engine
(LLM or non-LLM) and an exchange executor. Unlike conventional OMS limits that assume upstream strategy intent is trusted,
SAE treats upstream outputs as \emph{untrusted intent} in agentic settings (prompt/skill/narrative contamination),
and enforces non-bypassable invariants at the last mile where side effects occur.
This section specifies SAE as (i) a deployable system design, and (ii) a set of practical algorithms that map replay-computable state
to enforceable budgets and decisions, compatible with the Binance replay protocol in Section~\ref{sec:eval_binance}.

\subsection{System design and execution contract}
\label{sec:sae_system}

\paragraph{Middleware placement.}
SAE sits in the last-mile boundary:
\[
\text{Strategy Engine} \rightarrow \text{ExecutionRequest} \rightarrow \text{SAE Middleware} \rightarrow \text{Exchange Executor}.
\]
This placement ensures constraints are \emph{non-bypassable}: even if upstream intent is manipulated, the executor only receives SAE-approved actions.

\paragraph{Execution contract (API).}
SAE standardizes an explicit execution contract:

\begin{itemize}
  \item \textbf{ExecutionRequest} $r_t$:
  \texttt{symbol}, \texttt{venue}, \texttt{timestamp\_ms}, \texttt{intent}, \texttt{side},
  \texttt{requested\_notional}, \texttt{requested\_leverage}, \texttt{order\_type},
  \texttt{max\_slippage\_bps}, \texttt{strategy\_id}.
  \item \textbf{ExecutionContext} $c_t$:
  account state (equity, drawdown, positions, recent trades, margin ratio),
  market state (volatility, funding, liquidity proxy, regime),
  plus a first-class \textbf{trust state} $z_t$ (Section~\ref{sec:sae_trust_budget}),
  and optional narrative/anomaly signals.
  \item \textbf{ExecutionDecision} $d_t$:
  $\texttt{decision}\in\{\texttt{ALLOW},\texttt{LIMIT},\texttt{BLOCK}\}$
  plus enforced constraints (leverage cap, notional cap, order-rate cap, slippage cap, cooldown, staging plan),
  and audit fields (rule hit, budgets, state snapshot, timing).
\end{itemize}

\paragraph{Concrete example (request $\rightarrow$ SAE decision).}
\begin{quote}
\begin{verbatim}
# Strategy asks for an aggressive trade
ExecutionRequest = {
  "symbol": "BTCUSDT",
  "side": "LONG",
  "leverage": 5.0,
  "notional_fraction": 0.50,
  "slippage_bps": 80
}

# SAE returns a safer executable action
ExecutionDecision = {
  "decision": "LIMIT",
  "effective_leverage": 1.0,
  "effective_notional_fraction": 0.20,
  "effective_slippage_bps": 30,
  "cooldown_sec": 120,
  "reason": "extreme regime + low skill provenance"
}
\end{verbatim}
\end{quote}

This example shows the practical meaning of SAE. The strategy requests an aggressive trade
(5$\times$ leverage, 50\% notional allocation, and a wide slippage bound), but SAE does not simply accept or reject it.
Instead, it returns a \textsf{LIMIT} decision and converts the request into a safer executable action:
leverage is reduced to 1$\times$, notional is reduced to 20\% of the default budget, slippage tolerance is tightened,
and a cooldown is imposed. In this way, SAE preserves the intent to trade while enforcing survivability constraints at the execution boundary.

\paragraph{Modular components.}
Each SAE module is independently deployable and testable:

\begin{enumerate}
  \item \textbf{Trader-State Service}: outputs calibrated risk escalation probability $p_t \in [0,1]$.
  \item \textbf{Market Regime Detector}: classifies $\{\textsf{calm}, \textsf{volatile}, \textsf{extreme}\}$
  using volatility, funding extremes, and liquidity proxies.
  \item \textbf{Policy Engine}: YAML-defined mapping from (regime, risk, trust) to decision and budgets.
  \item \textbf{Enforcement Runtime}: last-mile projection/clamping, cooldown, rate limits, staged execution, and audit logging.
\end{enumerate}

Volatility regime detection is motivated by heteroskedasticity modeling; ARCH/GARCH formalize volatility clustering
\citep{engle1982arch,bollerslev1986garch}. In SAE we use deterministic replay-computable proxies rather than fitting heavy time-series models in the hot path.

\subsection{State construction from Binance replay (reproducible)}
\label{sec:sae_state}

SAE must run identically in offline replay and production. We therefore define a per-step state that can be computed deterministically from
Binance USD-M futures replay streams (15m bars, funding history).

\paragraph{Market features.}
Let $p_t$ be the close price and $r_t=\log p_t-\log p_{t-1}$.
Realized volatility over a rolling window $W_{\sigma}$:
\begin{equation}
\sigma_t=\sqrt{\sum_{i=t-W_{\sigma}+1}^{t} r_i^2 }.
\label{eq:sae_vol}
\end{equation}
Funding rate $f_t$ is aligned to the bar timeline (merged from funding history).
A simple deterministic liquidity proxy (any fixed proxy is acceptable as long as it is consistent across runs):
\begin{equation}
\lambda_t=\frac{\mathrm{EMA}(v_t)}{\sigma_t+\epsilon},
\label{eq:sae_liq}
\end{equation}
where $v_t$ is volume and $\epsilon>0$ avoids division by zero.

\paragraph{Account features.}
Account state includes equity $E_t$, peak equity $E_t^{\max}$, drawdown
$\mathrm{DD}_t = 1 - E_t/E_t^{\max}$, margin ratio $m_t$,
positions (direction, notional, leverage), and pacing statistics
(e.g., orders per window $W_{\text{rate}}$, average holding time, recent realized PnL).

\subsection{Intended-policy specification and out-of-scope rules (DG-ready)}
\label{sec:sae_intended_spec}

SAE adopts the Intended Policy Spec and hard out-of-scope rules defined in
Sections~\ref{sec:aintended_spec}--\ref{sec:oos_rules}. We use the same
$\mathcal{S}_t=(\mathcal{T}_t,\mathcal{R}_t,\mathcal{M}_t,\mathcal{U}_t)$
to (i) parameterize intended scope, (ii) enforce allowlists and caps at the executor boundary, and
(iii) label out-of-scope attempts deterministically for DG and attack instrumentation.

Concretely, $\mathcal{T}_t$ is implemented as allowlists over tools, venues, symbols, and accounts;
$\mathcal{R}_t$ becomes executable caps (leverage, notional, rate, slippage, holding time);
and $\mathcal{M}_t/\mathcal{U}_t$ become state predicates that trigger tightening or reduce-only modes in the enforcement runtime.
This alignment ensures that the same specification drives both enforcement and measurement.

\subsection{Trust state and trust-conditioned budgeting}
\label{sec:sae_trust_budget}

\paragraph{Trust state.}
Agentic execution adds a trust dimension beyond market/account state. SAE models trust as:
\begin{equation}
z_t=\big(p^{\mathrm{prov}}_t,\ r^{\mathrm{cap}}_t,\ \mathbb{I}^{\mathrm{inj}}_t\big),
\label{eq:trust_state}
\end{equation}
where $p^{\mathrm{prov}}_t\in[0,1]$ scores provenance of the active skill/toolchain,
$r^{\mathrm{cap}}_t$ scores capability risk (privilege and side-effect surface),
and $\mathbb{I}^{\mathrm{inj}}_t\in\{0,1\}$ is an injection-alert flag from safety monitors.

\paragraph{Budget vector and tightening.}
Let $B_t\in\mathbb{R}^k_{\ge 0}$ be a budget vector (e.g., leverage cap, notional cap, order-rate cap, slippage cap, max holding time).
We explicitly condition budgets on market/account/trust state:
\begin{equation}
B_t
=
B_0 \cdot g(\sigma_t,\lambda_t,f_t)\cdot h(m_t,\mathrm{DD}_t)\cdot q(z_t),
\label{eq:budget_factorized}
\end{equation}
where $B_0$ is the default budget,
$g(\cdot)$ tightens under volatile/illiquid/funding-extreme regimes,
$h(\cdot)$ tightens under constrained accounts (low margin / elevated drawdown),
and $q(\cdot)$ tightens when provenance is weak, capability risk is high, or injection alerts trigger.

\paragraph{Variant-to-switch mapping (for ablations).}
We align SAE variants in Section~\ref{sec:eval_variants_metrics} to algorithmic switches:
\begin{itemize}
\item \textsc{NoSAE}: bypass policy and enforcement (pass-through).
\item \textsc{Budget}: enable projection into $\mathcal{F}(B_t)$ using budgets from $B_0$ (optionally regime-conditioned $g$).
\item \textsc{Budget+Cooldown}: \textsc{Budget} + temporal invariants (cooldown, order-rate caps).
\item \textsc{Full}: \textsc{Budget+Cooldown} + trust-conditioned tightening $q(z_t)$ and additional policy checks (tool/state predicates).
\end{itemize}

\paragraph{Numerical walkthrough (trust tightening in one step).}
Consider a request with $L_{\mathrm{req}}=5\times$ leverage in an \textsc{extreme} regime, produced by a skill with low provenance
$p^{\mathrm{prov}}_t=0.3$ and an active injection alert $\mathbb{I}^{\mathrm{inj}}_t=1$.
Suppose the default leverage budget is $B_{0,L}=3\times$ and the default notional budget is $B_{0,N}=1.0$ (normalized).
Let regime tightening yield $g_L(\sigma_t,\lambda_t,f_t)=\tfrac{1}{3}$ and trust tightening yield $q_L(z_t)=0.5$.
Then the effective leverage cap becomes
\[
L_{\mathrm{cap}} = B_{0,L}\cdot g_L \cdot q_L = 3 \times \tfrac{1}{3} \times 0.5 = 0.5\times,
\]
which is clamped to $1\times$ in practice. If, for notional, $g_N=0.5$ and $q_N=0.4$, then
\[
N_{\mathrm{cap}} = B_{0,N}\cdot g_N \cdot q_N = 0.20,
\]
so the notional cap drops to $20\%$ of default. The returned decision is therefore \textsf{LIMIT}
with $(L_{\mathrm{cap}}=1\times,\ N_{\mathrm{cap}}=0.20)$, optionally with cooldown enabled under \textsc{extreme}.

\label{sec:sae_diagrams}

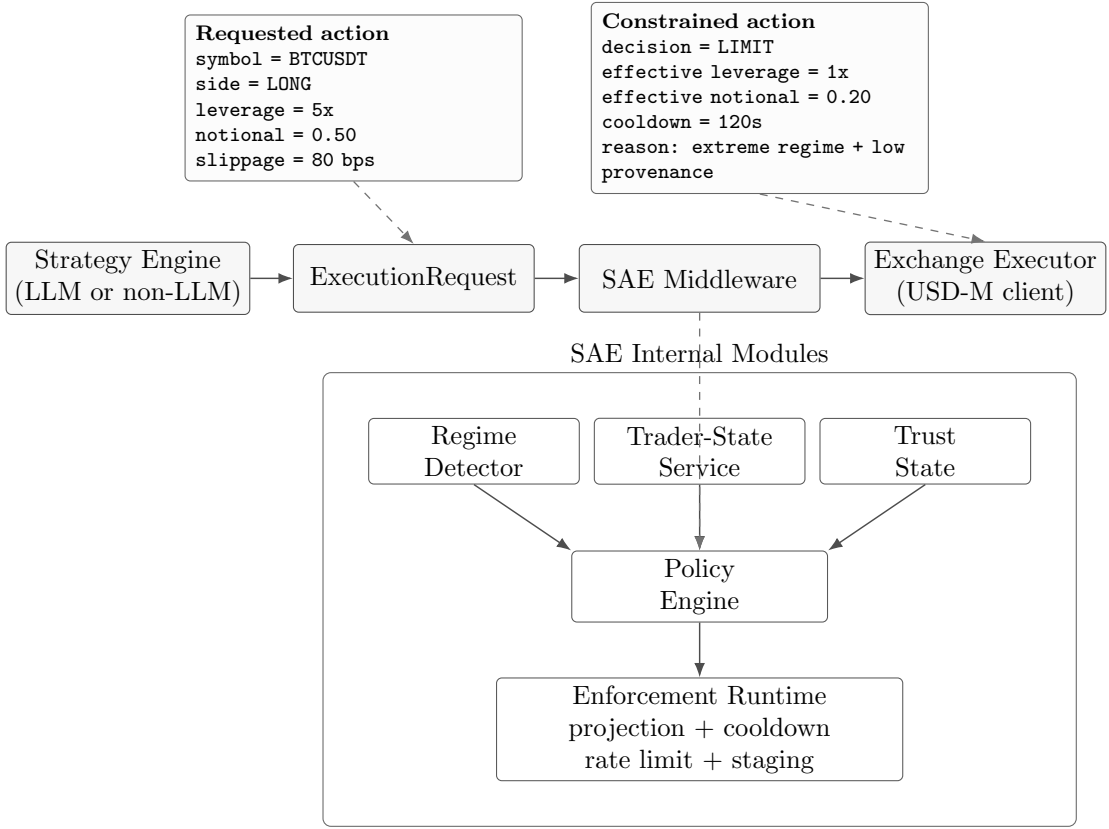
\begin{figure}[t]
\centering
\begin{tikzpicture}[
  font=\small,
  >=Latex,
  stage/.style={
    draw=black!70, rounded corners=2pt, align=center,
    minimum height=9mm, minimum width=32mm,
    inner sep=3pt, fill=black!3
  },
  mod/.style={
    draw=black!65, rounded corners=2pt, align=center,
    minimum height=8.5mm, minimum width=28mm,
    inner sep=3pt, fill=white
  },
  api/.style={
    draw=black!60, rounded corners=2pt, align=left,
    inner sep=4pt, fill=black!1, font=\scriptsize,
    text width=42mm
  },
  group/.style={
    draw=black!60, rounded corners=3pt, inner sep=6mm
  },
  arr/.style={-Latex, line width=0.55pt, draw=black!70},
  dasharr/.style={-Latex, dashed, line width=0.5pt, draw=black!55}
]

\node[stage] (strat) at (0,0) {Strategy Engine\\(LLM or non-LLM)};
\node[stage] (req)   at (3.8,0) {ExecutionRequest};
\node[stage] (sae)   at (7.6,0) {SAE Middleware};
\node[stage] (exec)  at (11.4,0) {Exchange Executor\\(USD-M client)};

\draw[arr] (strat) -- (req);
\draw[arr] (req) -- (sae);
\draw[arr] (sae) -- (exec);

\node[api] (reqex) at (3.0,2.4) {\textbf{Requested action}\\
\texttt{symbol = BTCUSDT}\\
\texttt{side = LONG}\\
\texttt{leverage = 5x}\\
\texttt{notional = 0.50}\\
\texttt{slippage = 80 bps}};

\node[api] (decex) at (8.4,2.4) {\textbf{Constrained action}\\
\texttt{decision = LIMIT}\\
\texttt{effective leverage = 1x}\\
\texttt{effective notional = 0.20}\\
\texttt{cooldown = 120s}\\
\texttt{reason: extreme regime + low provenance}};

\draw[dasharr] (reqex.south) -- (req.north);
\draw[dasharr] (decex.south) -- (exec.north);

\node[mod] (mr)    at (4.6,-2.3) {Regime\\Detector};
\node[mod] (ts)    at (7.6,-2.3) {Trader-State\\Service};
\node[mod] (trust) at (10.6,-2.3) {Trust\\State};

\node[mod, minimum width=34mm] (pe) at (7.6,-4.1) {Policy\\Engine};
\node[mod, minimum width=54mm] (enf) at (7.6,-6.0)
{Enforcement Runtime\\projection + cooldown\\rate limit + staging};

\node[group, fit=(mr)(ts)(trust)(pe)(enf),
      label={[font=\small]above:SAE Internal Modules}] (grp) {};

\draw[arr] (mr.south) -- (pe.north west);
\draw[arr] (ts.south) -- (pe.north);
\draw[arr] (trust.south) -- (pe.north east);
\draw[arr] (pe.south) -- (enf.north);

\draw[dasharr] (sae.south) -- (pe.north);

\end{tikzpicture}
\caption{SAE sits between strategy and executor as a non-bypassable execution layer. The top boxes show a concrete before/after example: an aggressive request is transformed into a constrained executable decision. The internal modules combine regime, trader-state, and trust signals into a policy decision, which is then enforced at the execution boundary.}
\label{fig:sae_overview}
\end{figure}

\label{sec:sae_pseudocode}

\begin{algorithm}[t]
\caption{Trader-state escalation predictor (calibrated)}
\label{alg:trader_state}
\begin{algorithmic}[1]
\Require Recent trades window $W$, account state (equity $E_t$, peak $E_t^{\max}$, margin ratio $m_t$), pacing stats, leverage-after-loss proxy
\State $x_t \leftarrow \mathrm{features}(W, E_t, E_t^{\max}, m_t)$
\State $\hat{p}_t \leftarrow \mathrm{base\_model}(x_t)$ \Comment{e.g., logistic regression or GBM}
\State $p_t \leftarrow \mathrm{calibrate}(\hat{p}_t)$ \Comment{isotonic or temperature scaling}
\State \Return $\mathrm{risk\_escalation\_score}=p_t$
\end{algorithmic}
\end{algorithm}

\begin{algorithm}[t]
\caption{YAML policy mapping to budgets and decision}
\label{alg:policy_engine}
\begin{algorithmic}[1]
\Require $\mathrm{regime}_t\in\{\textsf{calm},\textsf{volatile},\textsf{extreme}\}$, risk score $p_t$, trust state $z_t$, defaults $B_0$, rule list $\mathcal{R}$
\State $B_t \leftarrow B_0$
\State $B_t \leftarrow B_t \cdot g(\sigma_t,\lambda_t,f_t)\cdot h(m_t,\mathrm{DD}_t)\cdot q(z_t)$ \Comment{Eq.~\eqref{eq:budget_factorized}}
\For{\textbf{each} rule $r \in \mathcal{R}$}
  \If{$r.\mathrm{pred}(\mathrm{regime}_t,p_t,z_t)$ is \textbf{true}}
    \State $(\mathrm{decision},B_t)\leftarrow r.\mathrm{apply}(B_t)$
    \State \textbf{break}
  \EndIf
\EndFor
\State \Return $\mathrm{decision}\in\{\textsf{ALLOW},\textsf{LIMIT},\textsf{BLOCK}\}$ and budgets $B_t$
\end{algorithmic}
\end{algorithm}

\begin{algorithm}[t]
\caption{Enforcement runtime as projection with operational guards}
\label{alg:enforcement}
\begin{algorithmic}[1]
\Require ExecutionRequest $a_{\text{req}}$, budgets $B_t$, decision, cooldown state, rate-limit state
\If{$\mathrm{decision}=\textsf{BLOCK}$}
  \State Reject (or output \textsc{NoOp}/reduce-only action); emit audit log with rule-hit and state snapshot
\Else
  \State $a_{\text{proj}} \leftarrow \arg\min_{a \in \mathrm{feasible}(B_t)} D(a,a_{\text{req}})$ \Comment{projection/clamping}
  \If{\textbf{violates} cooldown or order-rate caps}
    \State $a_{\text{eff}} \leftarrow \textsc{NoOp}$; emit audit log
  \Else
    \State $a_{\text{eff}} \leftarrow \mathrm{stage}(a_{\text{proj}})$ \Comment{sliced execution if enabled}
  \EndIf
  \State Enforce slippage bounds and venue/tool allowlists; emit audit log
\EndIf
\State \Return $a_{\text{eff}}$ and audit record
\end{algorithmic}
\end{algorithm}

\subsection{Projection-based enforcement: feasibility and a robust bound}
\label{sec:sae_projection_bound}

\paragraph{Feasible set and projection.}
Let $u(a)\in\mathbb{R}^d$ be a normalized action representation (e.g., leverage, notional, order-rate, slippage tolerance).
Budgets induce a feasible set $\mathcal{F}(B_t)$.
SAE computes an effective action by projection:
\begin{equation}
a_t^{\mathrm{eff}}=\arg\min_{a\in\mathcal{F}(B_t)} D(a,a_t^{\mathrm{req}}),
\label{eq:sae_projection}
\end{equation}
where $D$ is an execution-distance functional (e.g., weighted $\ell_2$ over normalized degrees of freedom).
A special case recovers leverage clamping:
\begin{equation}
L_{\mathrm{eff}}=\min(L_{\mathrm{req}},L_{\mathrm{cap}}).
\end{equation}

\paragraph{A weak but robust one-step loss bound.}
Let $e_t\in\mathbb{R}^d$ denote the exposure vector induced by an effective action (notional, leverage, rate, slippage budget, etc.).
Assume the per-step loss proxy $\ell_t(e)$ is $L$-Lipschitz w.r.t.\ a chosen norm:
$|\ell_t(e)-\ell_t(e')|\le L\|e-e'\|$.
If SAE enforces $\|e_t\|\le B_t$, then
\begin{equation}
\ell_t(e_t)-\ell_t(0)\le L\|e_t\|\le L B_t,
\end{equation}
and under projection $e_t^{\mathrm{eff}}=\Pi_{\mathcal{F}(B_t)}(e_t^{\mathrm{req}})$ we additionally have
\begin{equation}
\ell_t(e_t^{\mathrm{eff}})-\ell_t(e_t^{\mathrm{req}})
\le L\cdot \mathrm{dist}(e_t^{\mathrm{req}},\mathcal{F}(B_t)).
\end{equation}
This bound avoids assuming a fixed return path or linear PnL scaling, and directly links last-mile projection to reduced worst-case
instantaneous loss amplification. Liquidation and exchange-specific margin mechanics are evaluated empirically in replay.

\subsection{Algorithm hyperparameters (defaults) and auto-optimization hooks}
\label{sec:sae_hyperparams}

\begin{table}[t]
\centering
\caption{Default SAE hyperparameters (editable in config; tuned via constrained auto-optimization in binance mode).}
\label{tab:hyperparams}
\begin{tabular}{@{}llc@{}}
\toprule
Component & Hyperparameter & Default \\
\midrule
Trader-state & Trade window size $N$ & 200 \\
Trader-state & Calibration & isotonic \\
Trader-state & Thresholds $(\tau_{\textsf{LIMIT}},\tau_{\textsf{BLOCK}})$ & (0.50, 0.70) \\
Regime & Realized-vol window (bars) $W_{\sigma}$ & 60 \\
Regime & Regime thresholds $(\tau_{\sigma,1},\tau_{\sigma,2})$ & (1.0, 2.0) \\
Funding & Extreme funding threshold $\tau_f$ & 0.01 \\
Liquidity & Illiquidity threshold $\tau_\lambda$ & 0.0 (proxy-specific) \\
Enforcement & Baseline leverage cap & $3.0\times$ \\
Enforcement & Volatile leverage cap & $2.0\times$ \\
Enforcement & Extreme leverage cap & $1.0\times$ \\
Enforcement & Cooldown (volatile / extreme) & 60s / 120s \\
Enforcement & Rate-limit window $W_{\text{rate}}$ & 60s \\
Enforcement & Staging slices (volatile / extreme) & 4 / 5 \\
\bottomrule
\end{tabular}
\end{table}

\paragraph{Practical tuning with feasibility constraints.}
In binance mode, we tune SAE parameters under hard operational constraints (e.g., FalseBlock $\le 0.20$ and AttackSuccess $\le 0.80$)
using the auto-optimization protocol described in Section 6. This ensures the design is not merely conceptual:
hyperparameters are selected to satisfy deployability constraints while maximizing survivability objectives on real replay data.


\section{Threat Model and Mitigations}
\label{sec:threats}

SAE is motivated by a structural shift: when an agent is granted tool and executor privileges,
\emph{untrusted text becomes a control surface for monetized side effects}.
This section formalizes the threat model for agentic crypto execution and specifies which risks SAE
is designed to mitigate, how mitigations map to enforceable invariants, and how attacks are operationalized
as reproducible tests in our Binance replay evaluation.

\subsection{Threat model: assets, adversaries, and success conditions}
\label{sec:threat_model}

\paragraph{System boundary.}
We consider the pipeline:
\[
\text{Strategy (LLM or non-LLM)} \rightarrow \text{SAE middleware} \rightarrow \text{Local executor} \rightarrow \text{Exchange API}.
\]
The strategy and its inputs (prompts, skills, narratives, external content) are treated as \emph{potentially compromised}.
SAE is assumed to run inside the executor boundary where it can intercept and enforce all outbound execution calls.

\paragraph{Assets.}
We protect (i) account capital and margin safety (avoid liquidation and tail drawdowns),
(ii) execution integrity (prevent unauthorized tools/venues/symbols and out-of-scope actions),
(iii) availability (prevent order flooding and pathological churn that destabilizes the executor),
and (iv) audit integrity (retain sufficient evidence for post-incident analysis).

\paragraph{Adversaries.}
We model adversaries with realistic capabilities for agent ecosystems:
\begin{itemize}
\item \textbf{Prompt injection (direct/indirect).} Attacker controls some instruction channel or retrieved content that the agent treats as guidance
\citep{liu2024formalizing,yi2023benchmarking}.
\item \textbf{Skill / plugin supply-chain attacker.} Attacker publishes or updates an installable skill that influences tool calls and execution behavior
\citep{xu2026agentskills}. This includes compromised registries or malicious packages distributed via popular discovery surfaces.
\item \textbf{Narrative / instruction contagion.} Attacker injects action-inducing norms via agent social channels, causing rapid policy drift or unsafe
execution cascades \citep{jiang2026moltbook,manik2026openclaw_moltbook}.
\item \textbf{Execution-layer stress adversary.} Attacker does not need to ``steal funds'' directly; inducing extreme leverage, order flooding, or
slippage tolerance during stress can amplify tail losses in perpetual markets \citep{he2022perp_fundamentals}.
\end{itemize}

\paragraph{Assumptions (what SAE does \emph{not} solve).}
SAE does not assume it can prevent all compromise or perfectly identify malicious content.
Instead it assumes: (i) upstream intent may be wrong or adversarial; (ii) local execution must be constrained anyway;
(iii) exchange mechanics (funding, maintenance margin tiers) can amplify execution mistakes.
SAE is not a custody solution and does not replace key management, 2FA, or exchange-side controls.

\paragraph{Attack success conditions.}
An attack is considered successful if it produces an executed action that violates the intended spec
(Section~\ref{sec:sae_intended_spec}) or violates enforced execution constraints. Concretely, success includes:
\begin{itemize}
\item \textbf{Out-of-scope execution:} $a_t \notin A_{\mathrm{intended}}(\mathcal{S}_t)$ passes through to the executor (tool/venue/state/cap violation).
\item \textbf{Constraint evasion:} effective leverage/notional/order-rate/slippage exceeds SAE caps or violates cooldown/rate limits.
\item \textbf{Tail amplification:} statistically meaningful worsening of survivability outcomes (higher MDD/CVaR or liquidation proxy)
relative to policy-compliant baselines under the same replay stream.
\end{itemize}

\subsection{SAE mitigations as enforceable invariants}
\label{sec:sae_mitigations}

SAE mitigations are deliberately \emph{execution-centric}: they are implemented where side effects occur and therefore cannot be bypassed by upstream compromise.

\paragraph{(M1) Intended-policy spec and hard out-of-scope rules.}
SAE operationalizes ``what is allowed'' using the structured Intended Policy Spec
$\mathcal{S}_t=(\mathcal{T}_t,\mathcal{R}_t,\mathcal{M}_t,\mathcal{U}_t)$ (Section~\ref{sec:sae_intended_spec}).
This yields deterministic out-of-scope labeling and enables enforceable allowlists:
authorized tools, venues, symbols, accounts; and state-dependent constraints (reduce-only under stress).

\paragraph{(M2) Trust-state--conditioned tightening (agent-specific safety input).}
SAE explicitly models trust as $z_t=(p^{\mathrm{prov}}_t,r^{\mathrm{cap}}_t,\mathbb{I}^{\mathrm{inj}}_t)$
and tightens budgets using a factorized map:
\[
B_t = B_0 \cdot g(\text{market})\cdot h(\text{account})\cdot q(z_t),
\]
so that low provenance, high capability risk, or injection alerts reduce exposure budgets even if the strategy requests more.
This makes ``untrusted intent'' non-optional in the enforcement logic.

\paragraph{(M3) Projection-based enforcement of exposure budgets.}
Given a requested action $a_t^{\mathrm{req}}$, SAE enforces feasibility via projection
(Section~\ref{sec:sae_projection_bound}):
\[
a_t^{\mathrm{eff}} = \arg\min_{a\in\mathcal{F}(B_t)} D(a,a_t^{\mathrm{req}}),
\]
which generalizes heuristic clamping and ensures effective actions stay inside the budget region.
This blocks parameter escalation attacks without requiring semantic understanding of the prompt.

\paragraph{(M4) Temporal invariants: cooldown and order-rate limiting.}
Many practical failure modes are temporal (order flooding, churn, rapid flip under stress).
SAE enforces cooldown timers and rate caps as hard invariants at the executor boundary,
preventing high-frequency abuse even if upstream generates repeated actions.

\paragraph{(M5) Slippage bounds and staged execution.}
Under stress, excessive slippage tolerance can act as a ``permission'' to trade at any price.
SAE enforces slippage caps and (optionally) staged execution (slicing notional into time-spaced orders),
reducing microstructure-induced loss amplification.

\paragraph{(M6) Auditability and incident response.}
Each decision emits an audit record (matched rule, budgets, effective action, and state snapshot including trust signals).
This supports post-mortems and reproducibility: given the replay stream and config, the same decision trace is re-generated.

\paragraph{Mitigation coverage matrix.}
Table~\ref{tab:threat_mitigation_matrix} summarizes which SAE mechanisms address which threat classes.

\begin{table}[t]
\centering
\caption{Threats and SAE mitigation coverage (execution-centric invariants).}
\label{tab:threat_mitigation_matrix}
\begin{tabular}{@{}p{0.28\linewidth}p{0.67\linewidth}@{}}
\toprule
Threat class & SAE mitigations (non-bypassable) \\
\midrule
Prompt injection / untrusted instructions 
& (M1) intended-spec allowlists + hard OOS rules; (M3) projection into $\mathcal{F}(B_t)$; (M4) cooldown/rate limits; (M6) audit. \\
Skill / plugin supply chain \citep{xu2026agentskills}
& (M2) trust-state tightening ($p^{\mathrm{prov}}$); (M1) tool/venue allowlists; (M3) projection; (M6) audit for provenance. \\
Narrative contagion in agent networks \citep{jiang2026moltbook}
& (M2) narrative/injection flags (optional) + tightening; (M4) temporal invariants; (M1) state-dependent reduce-only modes. \\
Execution-layer stress (escalation, flooding, slippage abuse) 
& (M3) budget projection; (M4) cooldown/rate caps; (M5) slippage bounds + staging; (M6) traceable enforcement. \\
\bottomrule
\end{tabular}
\end{table}

\subsection{Attack suite for evaluation (behavioral safety tests)}
\label{sec:attack_suite}

To avoid triviality (e.g., only ``request 20$\times$ leverage''), we define a family of reproducible attack generators.
Each generator produces an \emph{out-of-scope action attempt stream} that is interleaved with the strategy’s normal intended actions.
All attacks are labeled using the deterministic out-of-scope rules induced by $\mathcal{S}_t$.

\paragraph{Attack families.}
We evaluate the following classes, parameterized for replay:
\begin{itemize}
\item \textbf{Parameter escalation:} attempt leverage/notional/slippage/order-rate values beyond $\mathcal{R}_t$ caps
(or beyond tightened caps under $\mathcal{M}_t/\mathcal{U}_t$).
\item \textbf{Cooldown bypass / order flooding:} attempt repeated rapid-fire actions to violate temporal invariants (M4).
\item \textbf{Tool/venue misuse:} attempt unauthorized tools, venues, symbols, or cross-account actions (violating $\mathcal{T}_t$).
\item \textbf{State-violation stress:} attempt risk-on actions while the system is in constrained states
(e.g., extreme volatility regime or low margin ratio where reduce-only is expected).
\item \textbf{Narrative-induced flip stress:} attempt rapid long/short flipping that is in-scope at the action-type level
but becomes out-of-scope under rate/cooldown and state-dependent tightening.
\end{itemize}

\paragraph{AttackSuccess, FalseBlock, and DG linkage.}
For an injected attack attempt stream $\{a_t^{\mathrm{atk}}\}$, we measure:
\begin{itemize}
\item \textbf{AttackSuccess (AS):} fraction of attack attempts that result in an executed action violating intended scope or enforced caps.
Operationally, AS counts an attempt as successful if the executor receives an effective action that is out-of-scope
or violates leverage/notional/rate/slippage/cooldown constraints.
\item \textbf{FalseBlock (FB):} fraction of legitimate in-scope intended actions that are blocked (or overly constrained if using a strict definition).
FB serves as an opportunity-cost proxy.
\item \textbf{DG metrics:} DG\_rate and DG\_loss are computed from the same out-of-scope labeling and loss proxies,
connecting attack robustness to measurable delegation-gap harm (Section~\ref{sec:sae_intended_spec}).
\end{itemize}

\paragraph{Reproducibility and feasibility constraints.}
All attack generators are seeded and replay-deterministic. In binance mode, we tune SAE hyperparameters under
feasibility constraints of the form:
\[
\mathrm{AS}(\pi)\le \alpha,\qquad \mathrm{FB}(\pi)\le \beta,\qquad \mathrm{Lat}(\pi)\le \tau,
\]
matching the constrained selection family in Eq.~(10) of our formulation.
In our reported auto-optimization protocol, we instantiate
$\beta=0.20$ and $\alpha=0.80$ (and report latency as an overhead metric), ensuring that improvements in survivability
do not come from trivially blocking everything.

\paragraph{Why this matters for perpetual markets.}
In perpetual futures, execution mistakes can be amplified via leverage and margin mechanics,
and funding cash flows can directly affect margin and liquidation risk. The attack suite therefore focuses on execution parameters that are structurally amplified (leverage, frequency, slippage)
and on state-triggered tightening where survivability is most fragile.

\section{Implementation and Reproducibility}
\label{sec:impl_repro}

\subsection{System Boundary and Microservice Interfaces}
\label{sec:impl_interfaces}

\paragraph{Design principle: non-bypassable last-mile enforcement.}
SAE is implemented as a \emph{policy-gated executor} that sits at the boundary where side effects occur (order placement,
cancellation, position modification). The core principle is that the LLM output is treated as \emph{untrusted intent}:
all proposed actions must pass a non-bypassable gating/projection step before they are executed.

\paragraph{Service decomposition.}
Our reference implementation uses a minimal, reproducible split:
(i) \textbf{MarketData} (Binance replay fetch + caching + alignment),
(ii) \textbf{AgentCore} (produces intended actions),
(iii) \textbf{SAE Gate} (budgeting, cooldown, trust-/state-conditioned tightening),
(iv) \textbf{Executor/Sim} (fills, fees, slippage, margin, liquidation),
and (v) \textbf{Logger} (equity curve, actions, safety events, DG labels).

\paragraph{Action schema.}
The agent produces an \texttt{IntendedAction} object; SAE maps it to \texttt{ALLOW}/\texttt{LIMIT}/\texttt{BLOCK}.
A minimal action schema is:
\begin{verbatim}
a_t = (type, symbol, side, notional, leverage, limit_px, meta)
type in {open, close, modify, cancel}
\end{verbatim}
The \texttt{meta} field carries tool provenance / trust flags (when available), enabling trust-conditioned budgeting.

\subsection{Binance Data Acquisition and Preprocessing}
\label{sec:impl_data}

\paragraph{Public endpoints (no API key required for market data).}
The replay dataset is constructed from Binance USD-M futures REST market endpoints:
\texttt{/fapi/v1/klines} (candlesticks),
\texttt{/fapi/v1/fundingRate} (funding history),
and \texttt{/fapi/v1/exchangeInfo} (contract metadata).
All fetched artifacts are cached on disk to ensure deterministic reruns.

\paragraph{Alignment and state construction.}
Klines define the master time index at the selected interval. Funding events are aligned to the nearest subsequent kline
boundary and merged into the per-step state. For each step we construct:
\[
s_t = (\mathrm{OHLCV}_t, \mathrm{funding}_t, \mathrm{spread/slip\_proxy}_t, \mathrm{account}_t, \mathrm{trust}_t).
\]
This state is consumed by both the agent and the SAE gate.

\subsection{Margin Accounting and Liquidation Engine}
\label{sec:impl_liquidation}

\paragraph{Margin and maintenance margin.}
We implement a configurable liquidation check. Liquidation is triggered when:
\[
\mathrm{margin\_balance}_t \le \mathrm{maintenance\_margin}_t.
\]
Maintenance margin is computed using the standard tiered form:
\[
\mathrm{maintenance\_margin} = \mathrm{notional}\cdot \mathrm{mmr} - \mathrm{maint\_amount},
\]
where $(\mathrm{mmr}, \mathrm{maint\_amount})$ are tier-dependent. If official tier tables are available, they can be supplied
as \texttt{configs/mm\_tiers\_<symbol>.csv}. If not, the system falls back to a conservative placeholder to avoid
overstating survivability.

\subsection{Auto-Optimization and Walk-Forward Parameter Selection}
\label{sec:impl_autoopt}

\paragraph{Constrained search with early stopping.}
We provide \texttt{auto\_optimize.py}, a batched best-so-far search around a checkpointed incumbent configuration.
Each sampled candidate is evaluated via a full replay and accepted only if it satisfies feasibility constraints:
\[
\mathrm{FalseBlock} \le \tau_{\mathrm{FB}}, \quad \mathrm{AttackSuccess} \le \tau_{\mathrm{AS}}.
\]
The search stops when no improvement is found for \texttt{patience} consecutive batches.

\paragraph{Reproducible commands and artifacts.}
The canonical command for Binance replay tuning is:
\begin{verbatim}
PYTHONPATH=. python scripts/auto_optimize.py \
  --mode binance \
  --config configs/default.yaml \
  --batch_trials 20 \
  --max_batches 15 \
  --patience 5 \
  --falseblock_max 0.20 \
  --attacksucc_max 0.80 \
  --resume
\end{verbatim}
The run writes:
\texttt{outputs\_auto/best.json} (best score + metadata),
\texttt{outputs\_auto/best\_full\_params.yaml} (best parameters),
and \texttt{outputs\_auto/final/<run\_id>/} (tables, figures, logs).
All experiments fix random seeds at the runner entry point and record configs to ensure bitwise reproducibility of outputs
under a fixed environment.


\subsection{Open Artifact: Installable SAE Policy Guard Skill (skills.sh)}
\label{sec:artifact_skills_sh}

\paragraph{Why a skills.sh artifact matters.}
Our threat model emphasizes that \emph{installable skill ecosystems} turn capability acquisition into a supply-chain problem:
an agent can extend its execution surface by installing third-party skills. To make SAE reproducible in such ecosystems,
we package SAE as an installable skills.sh skill, \texttt{sae-policy-guard}.\footnote{\url{https://skills.sh/true-ai-labs/sae-policy-guard/sae-policy-guard}}

\paragraph{What the artifact implements.}
The skill instantiates SAE as \emph{execution-layer middleware} that runs \textbf{before} any order reaches the exchange executor.
Given an intended action and a context snapshot (market state, account state, and optional trust signals), it returns a
three-way decision:
\[
\texttt{ALLOW} \;/\; \texttt{LIMIT} \;/\; \texttt{BLOCK},
\]
together with a projected risk budget (e.g., leverage/notional/rate/slippage caps) and a human-readable rationale string,
making the enforcement decision auditable.

\paragraph{Installation and minimal usage.}
The skill is installable via the skills CLI:
\begin{verbatim}
npx skills add https://skills.sh/true-ai-labs/sae-policy-guard/sae-policy-guard
\end{verbatim}
In our experiments, the policy guard is invoked as a pre-trade gate; the simulator logs both the \emph{requested} and
\emph{effective} (post-projection) actions for DG accounting and for reporting \texttt{AttackSuccess}/\texttt{FalseBlock}.

\paragraph{Reproducibility linkage.}
We recommend reporting (i) the exact skill version or commit hash, (ii) the exported policy spec (caps, cooldown rules),
and (iii) the tuning output (\texttt{best\_full\_params.yaml}). This makes the paper’s SAE evaluation
reproducible in a real skills-enabled agent stack.

\subsection{Reproducibility Checklist}
\label{sec:impl_checklist}

We recommend reporting the following items for reproducibility:
(i) symbols, interval, and time window (from \texttt{configs/default.yaml});
(ii) commit hash or version tag of the code package;
(iii) the best-params YAML exported by \texttt{auto\_optimize};
(iv) cached Binance replay files and their checksums; and
(v) the exact command line used for the reported tables/figures.

\section{Evaluation on Binance Replay}
\label{sec:eval_binance}

\subsection{Experimental Design}
\label{sec:eval_design}

\paragraph{Objective.}
We evaluate whether \textbf{Survivability-Aware Execution (SAE)} reduces tail-risk and execution-induced failure
modes when an agent is granted execution privileges. Our evaluation focuses on:
(i) drawdown and tail losses,
(ii) robustness to out-of-scope action attempts and delegation-gap violations, and
(iii) operational costs such as latency and false blocks.

\paragraph{Data: Binance USD-M futures replay.}
We replay \textbf{real Binance USD-M perpetual futures} market data for \textbf{BTCUSDT} and \textbf{ETHUSDT}
using a \textbf{15-minute} bar interval over \textbf{2025-09-01 to 2025-12-01}.
The dataset is constructed from Binance public REST endpoints (no API key required for market data),
including candlestick (klines), funding-rate history, and exchange metadata.
Funding rates are aligned to the kline timeline and merged into the per-step simulator state.

\paragraph{Backtest \& execution simulator.}
We use a simplified perpetual futures simulator with:
(i) taker fees and slippage modeled as a function of notional,
(ii) margin accounting with a configurable liquidation check (Section~\ref{sec:impl_liquidation}), and
(iii) an agent loop that generates intended actions which are gated by SAE before reaching the executor.
Each variant produces an equity curve and an action stream, from which we compute risk, safety, and usability metrics.

\subsection{SAE Variants and Measurements}
\label{sec:eval_variants_metrics}

\paragraph{Variants.}
We compare five variants:
\textsc{NoSAE} (no gating),
\textsc{StaticOMS} (fixed leverage cap + position-size limit + stop-loss + rate limit; no regime detection, no trust-state),
\textsc{Budget} (exposure budgeting / projection with state-based tightening but no trust-state),
\textsc{Budget+Cooldown} (budgeting plus cooldown/order-rate constraints),
and \textsc{Full} (budgeting plus trust-/state-conditioned tightening and additional policy checks).

\paragraph{Attack and Delegation-Gap instrumentation.}
We inject an adversarial stream of \emph{out-of-scope} action attempts (e.g., tool misuse, cap violations, or state violations,
depending on the configured attack generator). We measure:
\textbf{AttackSuccess} (fraction of attack attempts that pass through),
\textbf{FalseBlock} (fraction of in-scope legitimate actions blocked),
and \textbf{DG metrics} (delegation-gap rate and a loss proxy) computed from out-of-scope executions and their associated
loss contribution under the intended-policy specification.

\paragraph{Risk metrics.}
We report:
\textbf{MDD} (maximum drawdown),
\textbf{CVaR$_{0.95}$} and \textbf{CVaR$_{0.99}$} on per-step returns,
\textbf{Liquidation rate}, and \textbf{Latency} (estimated per-step decision overhead attributable to SAE).

\subsection{Results}
\label{sec:eval_results}

\paragraph{Main results.}
Table~\ref{tab:main_results_binance} summarizes risk and safety metrics on the Binance replay.
Adding \textsc{StaticOMS} makes clear that SAE adds value not merely on top of an unprotected executor, but also on top of conventional fixed-rule risk management.
Relative to \textsc{NoSAE}, all SAE variants dramatically reduce tail risk (CVaR) and drawdown while reducing the attack success rate.
Relative to \textsc{StaticOMS}, SAE further improves survivability by introducing state-aware and trust-aware execution constraints rather than relying only on fixed caps and stop-loss logic.
In this replay, \textsc{Budget} and \textsc{Budget+Cooldown} achieve the best risk profile among SAE variants, while \textsc{Full} incurs higher overhead and shows slightly worse drawdown/CVaR than budget-only variants, highlighting the importance of calibration and ablation-based design.

\begin{table}[H]
\centering
\caption{Binance replay (BTCUSDT, ETHUSDT; 15m; 2025-09-01--2025-12-01). Metrics aggregated over symbols. StaticOMS denotes a conventional fixed-rule risk layer (fixed leverage cap, stop-loss, rate limit, and position-size limit) without regime detection or trust-state conditioning.}
\label{tab:main_results_binance}
\setlength{\tabcolsep}{3.5pt}
\renewcommand{\arraystretch}{1.08}
\small
\begin{tabularx}{\linewidth}{lrrrrrrrrr}
\toprule
Variant & MDD & CVaR$_{0.95}$ & CVaR$_{0.99}$ & Liq. & AS & FB & Lat. (ms) & DG$_r$ & DG$_\ell$ \\
\midrule
NoSAE            & 0.4643 & $-1.363\times10^{-3}$ & $-4.025\times10^{-3}$ & 0 & 1.0000 & 0.0000 & 0.00135 & 0.05889 & 0.64712 \\
StaticOMS        & 0.1184 & $-2.34\times10^{-4}$  & $-5.62\times10^{-4}$  & 0 & 0.8847 & 0.0112 & 0.00194 & 0.05160 & 0.11890 \\
Budget           & 0.03251& $-5.10\times10^{-5}$  & $-1.01\times10^{-4}$  & 0 & 0.7600 & 0.0000 & 0.00251 & 0.04222 & 0.02132 \\
Budget+Cool  & 0.03627& $-5.90\times10^{-5}$  & $-1.03\times10^{-4}$  & 0 & 0.7588 & 0.0000 & 0.00321 & 0.04333 & 0.02553 \\
Full             & 0.03190& $-4.80\times10^{-5}$  & $-1.02\times10^{-4}$  & 0 & 0.7281 & 0.0000 & 0.01029 & 0.04700 & 0.01916 \\
\bottomrule
\end{tabularx}

\vspace{2mm}
\footnotesize
\textbf{Notes:} Liq.=liquidation count; AS=AttackSuccess; FB=FalseBlock; Lat.=decision overhead; DG$_r$=DG rate; DG$_\ell$=DG loss proxy. StaticOMS captures standard fixed-rule risk management without trust-state conditioning or regime-aware tightening.
\end{table}

\paragraph{Interpreting AttackSuccess under \textsc{Full}.}
The remaining AttackSuccess of $0.728$ indicates that \textsc{Full} substantially reduces, but does not eliminate, adversarial leakage.
In our attack suite, explicit cap violations and unauthorized tool/venue requests are usually caught by projection and allowlists.
The residual successes arise primarily from state-sensitive and timing-sensitive attempts that fall near regime boundaries or remain formally in-scope at the request level but are still harmful in context.
In particular, when the gate is calibrated to preserve zero FalseBlock in this run, ambiguous cases are more often \textsf{LIMIT}ed than \textsf{BLOCK}ed.
This should therefore be interpreted as a practical trade-off rather than a contradiction:
\textsc{Full} prioritizes survivability improvement with no observed false blocking, but still leaves room for stricter configurations that would further reduce AttackSuccess at the cost of higher intervention. While SAE substantially reduces \texttt{AttackSuccess} in our current evaluation (from 1.00 to 0.728), further lowering this metric---especially under stronger and adaptive attack suites---is an important direction for future work and will be addressed in a subsequent SAE v2 design.

\begin{figure}[t]
\centering
\includegraphics[width=0.75\linewidth]{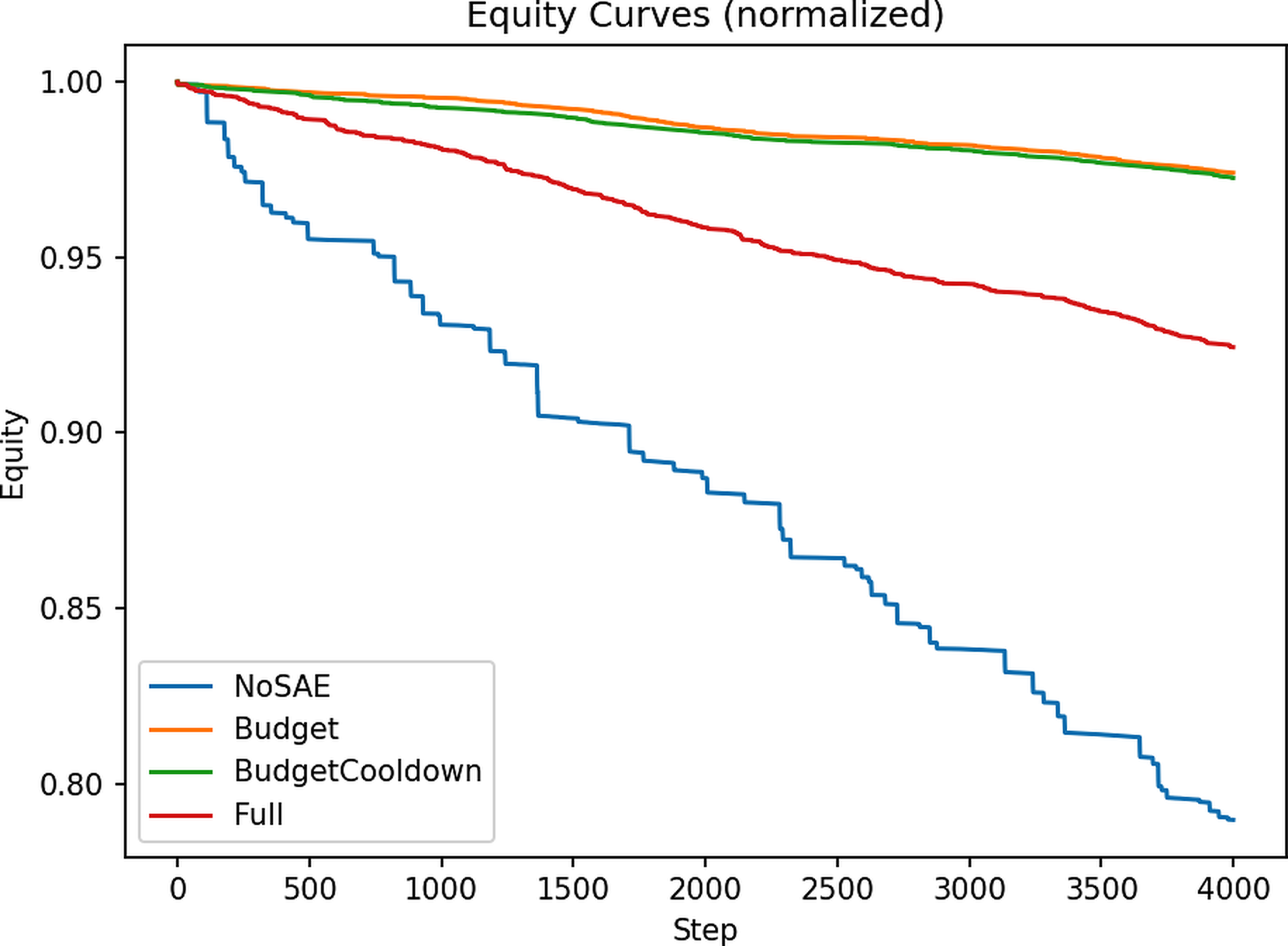}
\caption{SAE-focused zoom of normalized equity curves (Budget / Budget+Cooldown / Full) on Binance replay.
The three SAE variants have close nominal trajectories in this window, motivating tail-risk and robustness metrics as primary endpoints.}
\label{fig:equity_sae_zoom}
\end{figure}

\begin{figure}[t]
\centering
\includegraphics[width=0.95\linewidth]{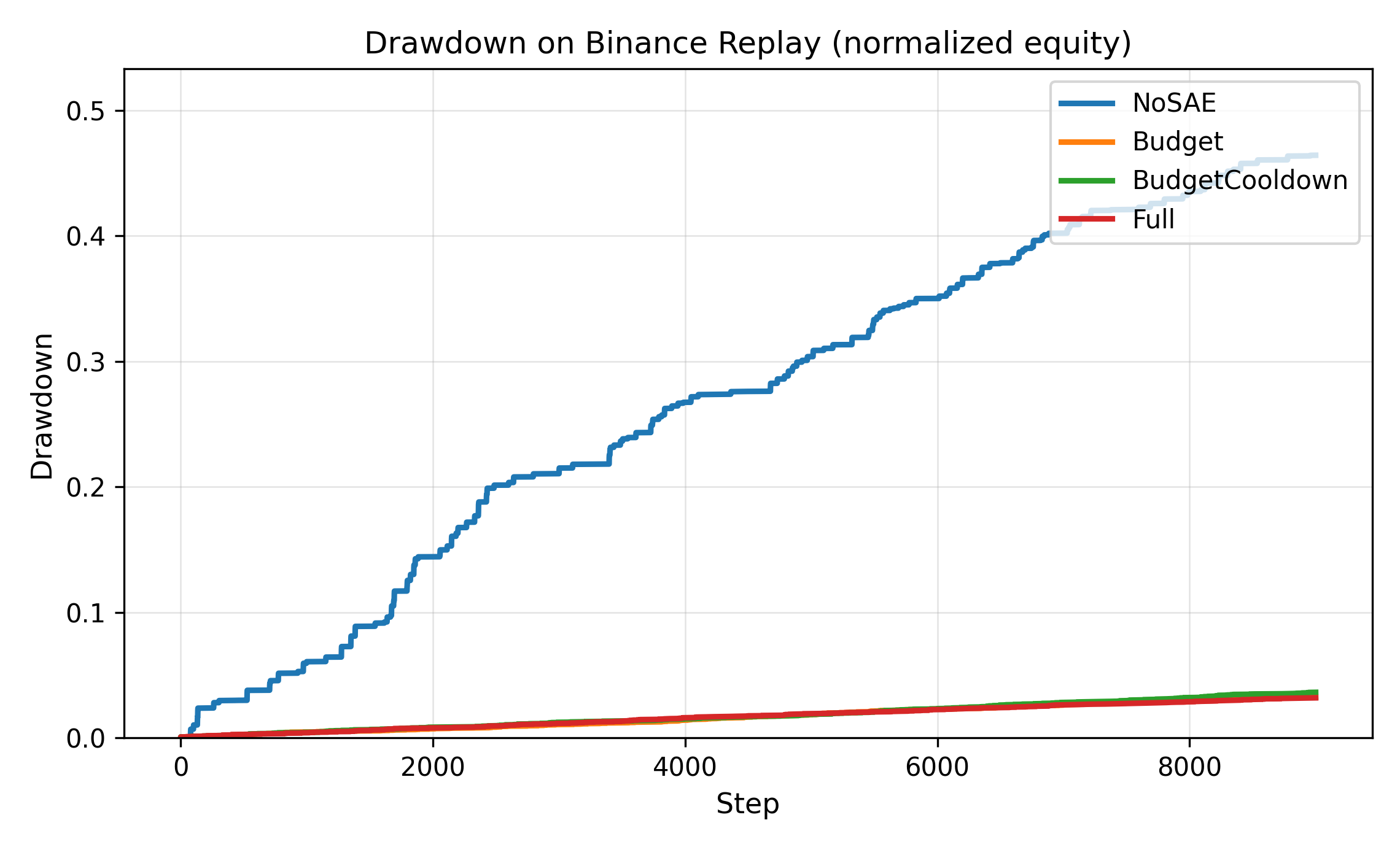}
\caption{Drawdown curves on Binance replay (BTCUSDT+ETHUSDT; 15m; 2025-09-01--2025-12-01). SAE variants sharply reduce maximum drawdown relative to \textsc{NoSAE}.}
\label{fig:drawdown_binance}
\end{figure}

\paragraph{Equity and drawdown.}
Figure~\ref{fig:equity_sae_zoom} highlights that the three SAE variants cluster closely in nominal regimes (SAE-focused zoom),
while Figure~\ref{fig:drawdown_binance} shows that \textsc{NoSAE} experiences substantially larger drawdowns driven by tail events.
This is consistent with SAE acting as a \emph{survivability layer} that primarily removes extreme downside rather than maximizing average returns.

\label{sec:eval_highlights}

Using Table~\ref{tab:main_results_binance}, SAE yields large survivability gains compared to \textsc{NoSAE}:
\begin{itemize}
    \item \textbf{Max drawdown (MDD):} \textsc{NoSAE} $0.4643 \rightarrow$ \textsc{Budget} $0.0325$ (\textbf{93.0\%} reduction);
    \textsc{NoSAE} $0.4643 \rightarrow$ \textsc{Full} $0.0319$ (\textbf{93.1\%} reduction).
    \item \textbf{Tail risk (CVaR$_{0.99}$):} tail-loss magnitude drops from $4.025\times10^{-3}$ to $\approx 1.02\times10^{-4}$
    (\textbf{$\sim$97.5\%} reduction in magnitude across SAE variants).
    \item \textbf{Delegation-gap harm:} DG\_loss decreases from $0.647$ to $0.019$--$0.026$
    (\textbf{96.1\%--97.0\%} reduction), indicating that out-of-scope executions no longer dominate loss contribution.
    \item \textbf{Attack robustness:} AttackSuccess decreases from $1.00$ to $0.728$ (\textbf{27.2 percentage-point} reduction) under \textsc{Full},
    while FalseBlock remains $0.00$ in this replay.
    \item \textbf{Operational overhead:} latency increases from $1.348\times10^{-3}$ ms (\textsc{NoSAE}) to $2.505\times10^{-3}$ ms (\textsc{Budget})
    and $1.029\times10^{-2}$ ms (\textsc{Full}), quantifying an enforcement-throughput trade-off (\textsc{Full} is $\sim 7.63\times$ \textsc{NoSAE} here).
\end{itemize}

\subsection{Uncertainty and Statistical Significance}
\label{sec:eval_significance}

\paragraph{Why robust tests matter.}
Per-step returns in market replay are temporally correlated; naive i.i.d.\ tests can overstate significance.
We therefore report dependence-aware uncertainty via \textbf{block bootstrap} confidence intervals for mean return,
a paired \textbf{Wilcoxon signed-rank test} on aligned per-step returns, and a \textbf{two-proportion z-test} for AttackSuccess.

\paragraph{Block bootstrap (mean return).}
The 95\% block-bootstrap confidence intervals for mean per-step return are:
\[
\textsc{NoSAE}: [-9.85,\,-5.01]\times10^{-5}, \qquad
\textsc{Full}: [-2.27,\,-1.71]\times10^{-5}.
\]
The intervals are well-separated, consistent with SAE removing catastrophic tail losses and improving the average return in this window.

\paragraph{Paired test on returns.}
A paired Wilcoxon signed-rank test comparing aligned per-step returns between \textsc{NoSAE} and \textsc{Full}
rejects the null of equal medians ($p = 0.0113$).

\paragraph{AttackSuccess significance.}
A two-proportion test for AttackSuccess between \textsc{NoSAE} and \textsc{Full} yields
$p = 1.76\times10^{-8}$, confirming that the reduction in out-of-scope attacks passing through is highly significant.

\begin{table}[t]
\centering
\caption{Significance summary on Binance replay (NoSAE vs Full).}
\label{tab:significance_binance}
\setlength{\tabcolsep}{6pt}
\begin{tabular}{lcc}
\toprule
Statistic & NoSAE & Full \\
\midrule
Mean return 95\% block-bootstrap CI & $[-9.85,\,-5.01]\times10^{-5}$ & $[-2.27,\,-1.71]\times10^{-5}$ \\
Wilcoxon signed-rank ($p$) & \multicolumn{2}{c}{$0.0113$} \\
Two-proportion test on AttackSuccess ($p$) & \multicolumn{2}{c}{$1.76\times10^{-8}$} \\
\bottomrule
\end{tabular}
\end{table}

\subsection{Discussion and Practical Takeaways}
\label{sec:eval_discussion}

\paragraph{Why ablations matter.}
Although SAE variants show similar nominal equity in this window (Figure~\ref{fig:equity_sae_zoom}), they differ in robustness and overhead.
Budget-only gating offers strong tail-risk reduction with low latency increase, while \textsc{Full} further reduces AttackSuccess and DG\_loss
but incurs higher latency. This motivates reporting ablations and explicitly characterizing the enforcement--throughput trade-off.

\paragraph{Limitations.}
Results depend on the configured attack generator, the intended-policy specification used for out-of-scope labeling,
and the replay window. Future evaluation should include regime-diverse windows (e.g., high-volatility weeks, cross-asset stress),
and report uncertainty intervals for tail-risk deltas and robustness rates under multiple attack distributions.

\section{Discussion}
\label{sec:discussion}

\paragraph{SAE as the missing contract in OpenClaw-style stacks.}
OpenClaw-style architectures already emphasize tool interception and enforceable boundaries.
SAE complements this by specifying a \emph{trading-domain execution contract} that can be enforced at the same gateway/executor boundary:
explicit context, auditable \textsf{ALLOW}/\textsf{LIMIT}/\textsf{BLOCK} decisions, and non-bypassable constraints on exposure, pace, and slippage.
This is precisely the layer that prompt-only safety and model alignment cannot reliably provide.

\paragraph{Why skills.sh makes ``untrusted intent'' unavoidable.}
Skill marketplaces such as skills.sh turn capability acquisition into an operational workflow.
From a security perspective, installing a skill is a supply-chain event: it can modify which tools are called, how parameters are chosen,
and how external content is trusted.
Therefore, upstream intent is not merely ``noisy''---it is structurally \emph{untrusted} in the presence of third-party skills.
SAE’s trust state and trust-conditioned tightening make this assumption explicit and enforceable.

\paragraph{Tail-risk gains without being an alpha model.}
SAE does not aim to improve prediction accuracy; it reduces the blast radius of execution mistakes.
The Binance replay shows that SAE variants often preserve nominal trajectories in typical regimes while producing large improvements in MDD/CVaR and DG loss.
This is consistent with SAE acting as a survivability layer: it prevents catastrophic actions (over-leverage, flood, excessive slippage, risk-on under stress) that dominate the tail.

\paragraph{Robustness must be reported as a systems metric.}
In the OpenClaw+skills setting, robustness is not a property of the model alone.
By operationalizing intent (Intended Policy Spec) and logging out-of-scope actions, SAE enables reproducible reporting of AttackSuccess, FalseBlock, DG rate, and DG loss.
This transforms qualitative security claims into measurable system-level metrics.

\paragraph{Ablations and the enforcement--throughput frontier.}
More checks are not always better.
Budget-only projection can deliver strong survivability improvements at low overhead, while trust-/state-conditioned tightening and extra policy checks
trade additional robustness for latency.
In practice, OpenClaw-style executors need a Pareto point that matches throughput requirements and acceptable adversarial leakage.
This motivates reporting ablations as first-class results and tuning SAE under explicit feasibility constraints.

\section{Limitations and Future Work}
\label{sec:limitations}

\paragraph{Venue mechanics and replay fidelity.}
Liquidation, maintenance margin tiers, fees, and market impact vary across venues and evolve over time.
Our replay uses a configurable simulator and aligned funding, but offline evaluation remains an approximation.
Future work should incorporate verified tier schedules per venue, richer impact models, and stress tests for liquidity holes.

\paragraph{Skill supply chain realism.}
While we model skill/provenance risk via a trust state, real ecosystems (e.g., skills.sh) introduce additional complexity:
versioning, dependency trees, transitive trust, and registry governance.
A next step is to integrate SBOM-style metadata, signatures, and reproducible build attestations into trust scoring,
and to define standardized provenance benchmarks for skill marketplaces.

\paragraph{Adaptive adversaries in OpenClaw-style tool interception.}
Attackers can probe gate behavior, adapt prompts, and route around policy checks by switching tools or venues.
Future work should evaluate adaptive multi-step adversaries, multi-venue execution, and cross-margin scenarios,
while keeping the evaluation protocol reproducible and safe.

\paragraph{Trust calibration and drift.}
Provenance scores, capability-risk scores, and injection alerts can be noisy and drift over time.
Miscalibration can increase FalseBlock or allow leakage.
Future work includes online calibration under drift, Bayesian trust aggregation, and learning-to-tighten policies that remain auditable.

\paragraph{Beyond trading: general-purpose executor safety.}
Although we focus on crypto execution, the OpenClaw+skills pattern applies broadly (payments, cloud ops, procurement).
A promising direction is a family of SAE-like domain contracts for privileged executors, plus a conformance suite:
``does this OpenClaw-style executor satisfy the contract and log sufficient evidence?''

\paragraph{Regime diversity and generalization.}
Our main results are for BTCUSDT/ETHUSDT over a specific window.
Future work should extend to regime-diverse periods (high-volatility weeks, cross-asset stress), additional symbols,
and uncertainty intervals across multiple seeds and attack distributions to quantify generalization.

\bibliographystyle{unsrtnat}
\bibliography{references}

\end{document}